\documentclass[numberedappendix]{emulateapj}
\usepackage{natbib}
\usepackage{graphicx}
\usepackage{multirow}

\def\sun{\ifmmode\odot\else$\odot$\fi}
\def\micron{\hbox{~$\mu$m}}

\newcommand{\Neii}{\hbox{[Ne\,{\sc ii}]12.81\micron}}
\newcommand{\Neiii}{\hbox{[Ne\,{\sc iii}]15.56\micron}}
\newcommand{\Siv}{\hbox{[S\,{\sc iv}]10.51\micron}}
\newcommand{\Neva}{\hbox{[Ne\,{\sc v}]14.32\micron}}

\newcommand{\Oiv}{\hbox{[O\,{\sc iv}]25.89\micron}}

\shorttitle{Local Luminous Infrared Galaxies. III. Co-evolution of Black
Hole Growth and Star Formation Activity?}
\shortauthors{Alonso-Herrero  et al.}

\begin{document}

\title{Local Luminous Infrared Galaxies. III.  Co-evolution of Black
Hole Growth and Star Formation Activity?}
\author{Almudena Alonso-Herrero\altaffilmark{1,2}, Miguel
  Pereira-Santaella\altaffilmark{3}, George
    H. Rieke\altaffilmark{4}, Aleksandar
M. Diamond-Stanic\altaffilmark{5}, Yiping Wang\altaffilmark{6},
Antonio Hern\'an-Caballero\altaffilmark{1} and Dimitra Rigopoulou\altaffilmark{7}}  
 \altaffiltext{$\star$}{This work is based on observations made with the Spitzer Space Telescope, which is operated by the Jet Propulsion Laboratory, California Institute of Technology under NASA contract 1407}
\altaffiltext{1}{Instituto de F\'{\i}sica de
  Cantabria, CSIC-Universidad de Cantabria, 39005 Santander, Spain} 
\altaffiltext{2}{Augusto Gonz\'alez Linares Senior Research Fellow}
\altaffiltext{3}{Istituto di Astrofisica e Planetologia Spaziali, INAF-IAPS, 
00133 Rome, Italy} 
\altaffiltext{4}{Steward Observatory, University
of Arizona, Tucson, AZ 85721, USA}
\altaffiltext{5}{Center for Astrophysics and Space Sciences, University of
California, San Diego, La Jolla, CA 92093, USA}
\altaffiltext{6}{National Astronomical Observatories, Chaoyang District
Beijing 100012,  China}
\altaffiltext{7}{Astrophysics Department, University
of Oxford, Oxford OX1 3RH, UK}

\begin{abstract}
Local  luminous infrared (IR) galaxies (LIRGs) have both 
high star formation rates (SFR) and a high AGN (Seyfert and
AGN/starburst composite) incidence. Therefore, 
they are ideal candidates to
explore the co-evolution of black hole (BH) growth and star formation (SF)
activity, not necessarily associated with major mergers.
Here, we use  {\it Spitzer}/IRS spectroscopy of 
a complete volume-limited sample of
local LIRGs (distances of $<78\,$Mpc). 
We estimate typical BH masses 
of $3\times 10^7\,M_\odot$ using  \Neiii \,and  optical [O\,{\sc
  iii}]$\lambda$5007 gas velocity dispersions and literature stellar velocity 
dispersions. We find that in a large
fraction of local LIRGs the current SFR is taking place not only in
the inner nuclear $\sim 1.5\,$kpc region, as estimated from 
the nuclear $11.3\,\mu$m PAH luminosities, but also in the
host galaxy. We next use the ratios between the SFRs and BH 
accretion rates (BHAR) to study whether the SF activity
and BH growth are contemporaneous in local LIRGs. 
On average, local LIRGs have SFR to BHAR  ratios higher 
than those of  optically selected Seyferts of similar AGN
luminosities. However, the majority of the
IR-bright galaxies in the RSA Seyfert sample behave like local
LIRGs. Moreover, the AGN incidence tends to be higher in local LIRGs
with the lowest SFRs. All this suggests that in local LIRGs there is 
a distinct IR-bright star forming phase taking place prior to the bulk
of the current BH growth (i.e., AGN phase). The
latter is reflected first as a composite and then as a Seyfert, and 
later as a non-LIRG optically
identified Seyfert nucleus with moderate SF in its host galaxy.

\end{abstract}

\section{Introduction}\label{s:intro}
One of the most fundamental relations in extragalactic astronomy is
that, at least in the local universe, the masses of  supermassive
black holes (BH, with masses $M_{\rm BH} > 10^{6}\,M_\odot$) correlate
with the stellar mass and velocity dispersion of the 
bulges of their host galaxies \cite[see e.g.][]{Magorrian1998,
  Gebhardt2000, Marconi2003, Haring2004}. This seems to imply that
bulges and supermassive BH evolve together and regulate each
other. 


Mergers of gas-rich galaxies are efficient in both producing elevated star
formation rates (SFR) and transporting gas to the nuclear region to
allow for BH
growth. If sufficient matter becomes available very close to the
nuclear BH and is accreted, the nucleus of the galaxy will shine with
enormous power as an active galactic nucleus \citep[AGN, ][]{Lynden-Bell1969}. 
In parallel, and probably before the fully developed AGN phase,
the merger will trigger a high rate of star formation leading to a
luminous infrared (IR)-active phase \citep{Sanders1988}, either as a luminous IR galaxy (LIRG)
or ultraluminous IR galaxy (ULIRG). These are defined as having an IR $8
-1000\,\mu$m luminosity $L_{\rm
  IR}>10^{11}\,L_\odot$ or $L_{\rm
  IR}>10^{12}\,L_\odot$, respectively \citep[see][for a review]{Sanders1996}.  

As an alternative to this sequence,  \cite{Kormendy2011} proposed that BHs in bulgeless galaxies and
in galaxies with pseudobulges grow through phases of {\it  low-level}
Seyfert-like activity. Such growth is believed to be 
driven stochastically by local processes (secular processes), and thus
it would not have a 
global impact on the host galaxy structure \citep[see
also][]{Hopkins2008}. Moreover, there is now 
evidence of two fundamentally different modes of BH growth at work in
early- and late-type galaxies  in
the local universe \citep[][]{Schawinski2010}. The role of
moderate luminosity AGN  that reside in late-type galaxies and do not involve
any recent major mergers is not clear and needs further investigation.

LIRGs are
powered by both AGN and star formation activity
\citep{Sanders1996}. 
\cite{AAH12} showed that while an AGN accompanies star
formation activity 
in a large proportion of local LIRGs, in most
cases the AGNs are not energetically important. This is because the
 AGNs hosted in local LIRGs have Seyfert-like luminosities and thus their
 bolometric contribution to the total IR luminosity is
 small (typically $5\%$). Therefore, the IR
luminosities of LIRGs imply SFRs in the range $11-110\,M_\odot \,{\rm
  yr}^{-1}$ using the
\cite{Kennicutt1998} prescription converted to  a 
\cite{Kroupa2002} Initial Mass Function (IMF). 

\begin{table}
\center
\caption{Spitzer/IRS Calibration sources }\label{table:AOR}
\begin{tabular}{lc}
\hline
\hline
Name & AOR\\
\hline
PNG043.1+37.7 & 4109056\\
PNG011.7-00.6 & 4109568\\ 
PNG025.8-17.9 & 4110080\\ 
PNG265.7+04.1 & 4111104\\ 
PNG206.4-40.5 & 4111616\\
PNG285.7-14.9 & 4112128\\ 
PNG054.1-12.1 & 4112640\\ 
PNG009.4-05.0 & 4114176\\
PNG342.1+10.8 & 4115200\\
PNG316.1+08.4 & 4116224\\ 
Calwav-22B-NGC7027 & 15343104\\ 
IRSS-SEPN-N7293-0001 & 15752960\\
\hline
\end{tabular}
\end{table}

It is reasonable to assume that the same gas that is used to form
stars in the host galaxy 
can also be used to feed the AGN provided that there is a
mechanism able to transport the gas to the inner region (on scales of
less than 0.1\,pc) of the galaxy \citep[see e.g. the review
of][]{Alexander2012}.  
Indeed, it is now apparent that there is a relation between star 
formation activity on sub-kiloparsec and bulge scales and the BH
accretion rate (BHAR), as shown
by observations and numerical 
simulations \citep[][]{Heckman2004, Hopkins2010,
  DiamondStanic2012}. This relation  
is also predicted to be present, although with a lower significance, for the
integrated SFR of the 
galaxy \citep{Hopkins2010}. Therefore, LIRGs show  propitious
conditions to study the   co-evolution of star formation activity and
 BH growth  as they show both 
  high integrated and nuclear SFRs \citep{AAH06} and a high
 occurrence of AGN and composite nuclei
 \citep[$60-70$\%,][]{Yuan2010,AAH12}.

In this paper we study the
co-evolution of the SFR activity and the BH growth in 
the complete volume-limited sample of local LIRGs of \cite{AAH06,
  AAH12}. We use {\it Spitzer} Infrared Spectrograph \cite[IRS, ][]{Houck2004}
observations to estimate the nuclear ($\sim 1.5\,$kpc) SFRs and
compare them with the integrated values from the IR luminosities. We
look for spectroscopically resolved \Neiii \, lines. Assuming that
this line is produced in the narrow line region (NLR) of the AGN, we 
measure its velocity dispersion and use the
\cite{Dasyra2011} relations to obtain the masses of the BH
hosted in local LIRGs. We complement these  with observations of the
optical line [O\,{\sc 
  iii}]$\lambda$5007 and literature values of the stellar velocity
dispersion $\sigma_{\rm *}$ to obtain further estimates of the BH masses in local
LIRGs. All this information allows us to determine the ratios
between the SFR (both, nuclear and integrated) and the BHAR 
in local LIRGs and compare them with those of optically
selected Seyfert galaxies. This comparison lets us explore the role of
the LIRG-phase in 
the growth of BH in the local universe. 
Throughout this paper we assume the following cosmology $H_0 = 70$ km
s$^{-1}$Mpc$^{-1}$, $\Omega_M = 0.3$ and $\Omega_{\Lambda} = 0.7$.

\section{Sample, Observations, and 
Data Reduction}\label{s:observations}

\subsection{The sample of local LIRGs}
For this work we use the complete volume-limited
sample of local LIRGs that was 
 presented and discussed in detail  by \cite{AAH06,AAH12}. We drew
 the sample  from the {\it IRAS} Revised Bright Galaxy Sample
 \citep[RBGS, ][]{Sanders2003} to include all the sources with
 $\log (L_{\rm IR}/L_\odot) \geq 11.05$ 
and $v_{\rm hel} =$ 2750 -- 5300 km s$^{-1}$. 
The sample is composed of 45 {\it IRAS} systems, with 8 
containing multiple nuclei. Therefore the sample includes 53
individual galaxies. For the assumed cosmology
the distances are in the range $\simeq 40-78\,$Mpc, with a median
value of 65\,Mpc. The IR luminosities of the individual
galaxies\footnote{Galaxies 
  significantly below the selection luminosity are members of multiple
  systems with total luminosities that satisfy the selection
  criterion.}  are in
the range  $\log (L_{\rm IR}/L_\odot) = 10.64-11.67$, with a median
value of  $\log (L_{\rm IR}/L_\odot) =11.12$. 
All the relevant information for the sample can
be found in table~1 of \cite{AAH12}.

Combining a number of optical and mid-infrared (mid-IR) indicators, \cite{AAH12}
derived an AGN detection rate of $\sim$62\% for this complete
volume-limited sample of local LIRGs. The derived AGN bolometric
luminosities,  from the mid-IR spectral decomposition and/or
 X-ray observations, are in the range $L_{\rm bol}({\rm AGN}) =
(0.4-50)\times 
10^{43}\,{\rm erg\,s}^{-1}$  with a median of 
$\sim 1.4 \times 10^{43}\,{\rm erg \,s}^{-1}$ \citep[][]{Pereira2011, AAH12}. However,
these AGN are  overall only responsible for $\sim 5\%$ of the IR luminosity
emitted by local LIRGs, although the AGN contribution varies from
source to source \cite[see][]{Imanishi2010, AAH12}. This is in good
agreement with \cite{Petric2011} 
for The Great Observatories All-Sky LIRG Survey \citep[GOALS;][]{Armus2009}.

\begin{figure}
\vspace{0.5cm}
\includegraphics[width=0.29\textwidth,angle=-90]{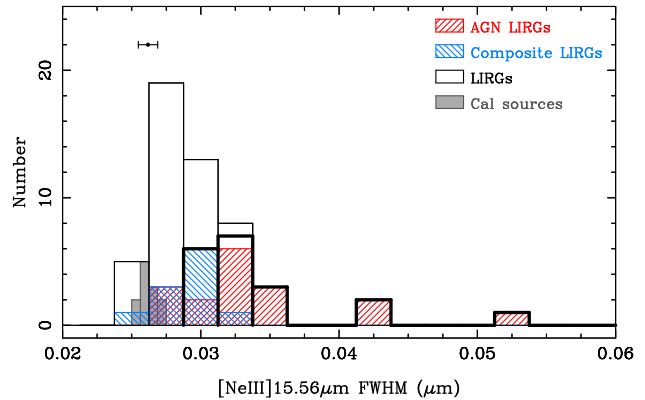}
\caption{Distribution of the observed (not corrected for instrumental
  resolution)  FWHMs of the \Neiii \, line of local LIRGs compared with those of the
  calibration sources. The latter are  assumed to  provide the instrumental
  resolution at the wavelength of the line. The dot represents
    the average value of the 
    FWHM of the calibration sources 
    with the corresponding $1\sigma$ error. We marked those
LIRG nuclei classified as AGN (Seyfert and/or [Ne\,{\sc v}] emitters) 
and composites. The rest are H\,{\sc
ii}-like or have no classification. The thick line histogram
represents the observed FWHM of those galaxies deemed to have
spectroscopically resolved lines.}
\vspace{0.5cm}
\label{fig:FWHMdistribution}
\end{figure}

\subsection{Spitzer/IRS Observations}
\subsubsection{Observations}\label{sec:calsources}
We retrieved {\it Spitzer}/IRS \citep{Houck2004}
spectroscopy of the sample of LIRGs taken with the high
resolution ($R\sim 600$) short-high (SH) and long-high (LH) modules 
that cover the $9.9-19.6\,\mu$m and $18.7-37.2\,\mu$m spectral ranges,
respectively. Details on the programme IDs, 
the observations, and data reduction of the galaxies, including those from
 GOALS \citep[][]{Armus2009}, are given in
\cite{AAH12} and \cite{Pereira2010IRSmapping, Pereira2010lines}. We
note that for various reasons two galaxies in this volume-limited
sample were  not observed with the {\it Spitzer}/IRS  high
spectral resolution modules.  We  extracted the nuclear spectra 
  assuming a point source calibration.

We also retrieved SH and LH spectroscopy from the {\it Spitzer} archive of
12 calibration sources and planetary nebulae to obtain an accurate estimate of the spectral
resolution using fine structure lines (see
Section~\ref{sec:instrumentalres}). The names and corresponding
Astronomical Observation Request (AOR) 
numbers are give in Table~\ref{table:AOR}. We reduced the data
as for the sample of local LIRGs.

\begin{figure*}
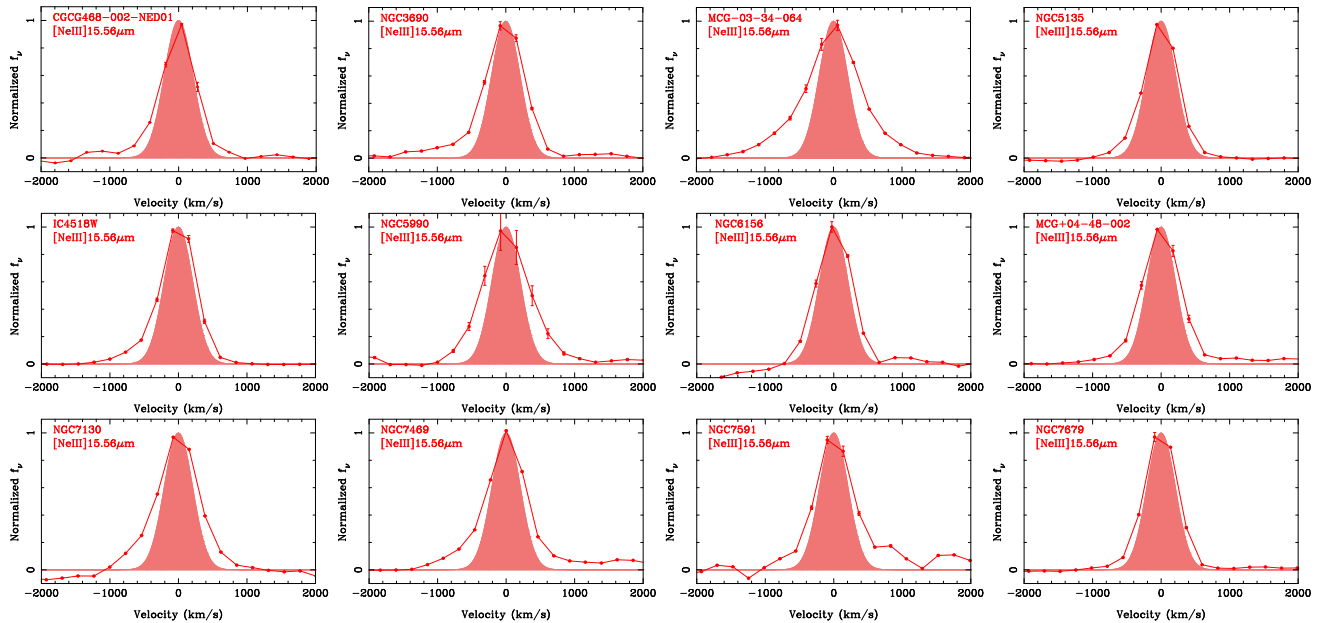

\center
\vspace{-0.1cm}
\includegraphics[width=0.15\textwidth,angle=-90]{figure2a.ps}
\includegraphics[width=0.15\textwidth,angle=-90]{figure2b.ps}
\includegraphics[width=0.15\textwidth,angle=-90]{figure2c.ps}
\includegraphics[width=0.15\textwidth,angle=-90]{figure2d.ps}
\includegraphics[width=0.15\textwidth,angle=-90]{figure2e.ps}
\includegraphics[width=0.15\textwidth,angle=-90]{figure2f.ps}
\includegraphics[width=0.15\textwidth,angle=-90]{figure2g.ps}
\includegraphics[width=0.15\textwidth,angle=-90]{figure2h.ps}
\includegraphics[width=0.15\textwidth,angle=-90]{figure2i.ps}
\includegraphics[width=0.15\textwidth,angle=-90]{figure2j.ps}
\includegraphics[width=0.15\textwidth,angle=-90]{figure2k.ps}
\includegraphics[width=0.15\textwidth,angle=-90]{figure2l.ps}

\caption{Observed profiles (solid lines and filled dots) 
of the spectrally resolved \Neiii \, lines of local
LIRGs that are classified as a Seyfert and/or are [Ne\,{\sc v}]
emitters. The shaded area shows a  
{\it Spitzer}/IRS SH unresolved profile
  represented as a Gaussian with  FWHM$=0.02618\,\mu$m, as 
determined from {\it Spitzer}/IRS calibration sources
  (Section~\ref{sec:calsources}).}
\label{fig:profiles}
\end{figure*}

\begin{figure*}
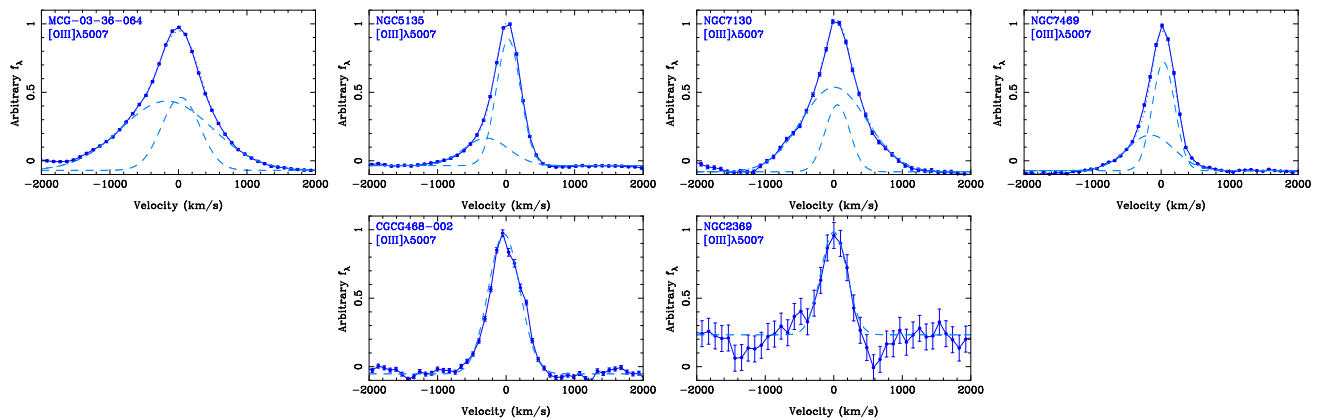

\center
\vspace{-0.1cm}
\includegraphics[width=0.15\textwidth,angle=-90]{figure3a.ps}
\includegraphics[width=0.15\textwidth,angle=-90]{figure3b.ps}
\includegraphics[width=0.15\textwidth,angle=-90]{figure3c.ps}
\includegraphics[width=0.15\textwidth,angle=-90]{figure3d.ps}
\includegraphics[width=0.15\textwidth,angle=-90]{figure3e.ps}
\includegraphics[width=0.15\textwidth,angle=-90]{figure3f.ps}
\caption{{\bf Top panel:} Observed profiles (solid lines and filled dots) of the [O\,{\sc
    iii}]$\lambda$5007 line 
  fitted with two Gaussian components (dashed lines) as explained in
  Section~\ref{sec:oiiiline}. {\bf Bottom panel: Same as top panel
  but for galaxies whose profiles are fitted with one Gaussian. The
  plotted errors for the 6dF data (NGC~2369, MGC~$-$03-34-064, NGC~5135, and
  NGC~7130) are those computed from the rms of the continuum adjacent to the
line.} }  
\label{fig:profilesOIII}
\end{figure*}

\subsubsection{Line Fitting}
For the fine-structure lines
we used Gaussians to  measure the line flux, equivalent width (EW), and full
width half maximum (FWHM) and a first order polynomial to fit the local
continuum \citep[see][for more details]{Pereira2010IRSmapping}. 
  The fluxes of the \Neii, \Neiii, and \Oiv \, lines can be found in 
\cite{Pereira2010lines} and \cite{AAH12}. For
the $11.3\,\mu$m polycyclic  
aromatic hydrocarbon (PAH) feature first we fitted the local continuum
using a linear fit between 10.6 and $11.8\,\mu$m. After subtracting
the continuum, we integrated 
the flux in the wavelength range between 10.8 and $11.6\,\mu$m. We
list the fluxes, EW, and corresponding errors of the $11.3\,\mu$m PAH
feature in Table~\ref{table:PAH}.

\begin{table}
\center
\caption{Fluxes and EW of the nuclear $11.3\,\mu$m PAH feature}\label{table:PAH}
\begin{tabular}{lcc}
\hline
\hline
Galaxy & Flux & EW\\
\hline
NGC~23 & 430.4 $\pm$ 8.5 & 0.78 $\pm$ 0.05 \\
MCG+12-02-001 & 396.8 $\pm$ 7.9 & 0.42 $\pm$ 0.01 \\
NGC~633 & 166.1 $\pm$ 4.0 & 0.61 $\pm$ 0.04 \\
ESO~297-G012 & 128.4 $\pm$ 3.5 & 0.60 $\pm$ 0.05 \\
UGC~01845 & 347.5 $\pm$ 5.2 & 0.74 $\pm$ 0.03 \\
UGC~02982 & 379 $\pm$ 35 & 0.84 $\pm$ 0.15 \\
CGCG~468-002 NED01 & 59.6 $\pm$ 5.6 & 0.24 $\pm$ 0.03 \\
CGCG~468-002 NED02 & 78.4 $\pm$ 7.7 & 0.46 $\pm$ 0.07 \\
UGC~03351 & 275.5 $\pm$ 8.3 & 0.78 $\pm$ 0.06 \\
NGC~2369 & 357.1 $\pm$ 3.9 & 0.58 $\pm$ 0.02 \\
NGC~2388 & 428 $\pm$ 14 & 0.53 $\pm$ 0.04 \\
MCG+02-20-003 & 149.5 $\pm$ 4.1 & 0.70 $\pm$ 0.06 \\
NGC~3110 & 299.0 $\pm$ 4.0 & 0.80 $\pm$ 0.03 \\
NGC~3256 & 1547.2 $\pm$ 4.8 & 0.556 $\pm$ 0.004 \\
ESO~264-G057 & 165 $\pm$ 13 & 0.76 $\pm$ 0.11 \\
IC~694 & 446 $\pm$ 16 & 0.38 $\pm$ 0.02 \\
NGC~3690 & 308 $\pm$ 25 & 0.09 $\pm$ 0.01 \\
ESO~320-G030 & 450.3 $\pm$ 4.6 & 0.78 $\pm$ 0.03 \\
MCG-02-33-098 W & 99.0 $\pm$ 4.2 & 0.27 $\pm$ 0.02 \\
MCG-02-33-098 E & 102.3 $\pm$ 6.8 & 0.62 $\pm$ 0.12 \\
IC~860 & 43.2 $\pm$ 2.2 & 0.44 $\pm$ 0.04 \\
MCG-03-34-064 & 23.0 $\pm$ 3.2 & 0.013 $\pm$ 0.002 \\
NGC~5135 & 376.9 $\pm$ 4.0 & 0.46 $\pm$ 0.01 \\
ESO~173-G015 & 434 $\pm$ 31 & 0.53 $\pm$ 0.06 \\
IC~4280 & 249.4 $\pm$ 7.0 & 0.80 $\pm$ 0.04 \\
UGC~08739 & 146 $\pm$ 41 & 0.84 $\pm$ 0.38 \\
ESO~221-IG010 & 259.9 $\pm$ 9.4 & 0.63 $\pm$ 0.05 \\
NGC~5653 & 385 $\pm$ 10 & 0.81 $\pm$ 0.07 \\
NGC~5734 & 402 $\pm$ 11 & 0.80 $\pm$ 0.06 \\
NGC~5743 & 251 $\pm$ 14 & 0.75 $\pm$ 0.12 \\
IC~4518 E & 64.6 $\pm$ 1.4 & 0.62 $\pm$ 0.05 \\
IC~4518 W & 62.6 $\pm$ 5.9 & 0.16 $\pm$ 0.02 \\
Zw~049.057 & 62.3 $\pm$ 4.5 & 0.96 $\pm$ 0.23 \\
NGC~5936 & 240.0 $\pm$ 6.6 & 0.67 $\pm$ 0.05 \\
NGC~5990 & 243 $\pm$ 56 & 0.29 $\pm$ 0.09 \\
NGC~6156 & 141 $\pm$ 16 & 0.22 $\pm$ 0.03 \\
IRAS 17138-1017 & 307.6 $\pm$ 4.3 & 0.53 $\pm$ 0.02 \\
IRAS 17578-0400 & 177 $\pm$ 10 & 0.76 $\pm$ 0.09 \\
IC~4687 & 478.2 $\pm$ 4.2 & 0.70 $\pm$ 0.02 \\
IC~4734 & 195.1 $\pm$ 3.8 & 0.59 $\pm$ 0.03 \\
NGC~6701 & 242.2 $\pm$ 4.2 & 0.67 $\pm$ 0.03 \\
MCG~+04-48-002 & 394 $\pm$ 33 & 0.73 $\pm$ 0.11 \\
NGC~7130 & 230.5 $\pm$ 3.7 & 0.36 $\pm$ 0.01 \\
IC~5179 & 391.9 $\pm$ 3.6 & 0.68 $\pm$ 0.02 \\
NGC~7469 & 583.5 $\pm$ 7.7 & 0.237 $\pm$ 0.005 \\
NGC~7591 & 142 $\pm$ 12 & 0.36 $\pm$ 0.06 \\
NGC~7679 & 363 $\pm$ 16 & 0.61 $\pm$ 0.05 \\
NGC~7769 & 85 $\pm$ 13 & 0.57 $\pm$ 0.21 \\
NGC~7770 & 131 $\pm$ 18 & 0.47 $\pm$ 0.13 \\
NGC~7771 & 359 $\pm$ 16 & 0.65 $\pm$ 0.07 \\

\hline
\end{tabular}

{\bf Notes.} The fluxes are in units of $10^{-14}\,{\rm erg\,
    cm}^{-2}\,{\rm s}^{-1}$ and the EW are in units of $\mu$m.
\end{table}

\subsubsection{Spectrally resolved \Neiii \, lines}\label{sec:instrumentalres}
The velocity dispersion of the ionized gas in the NLR allows measuring
the BH mass ($M_{\rm BH}$) in an AGN supplementing reverberation mapping
techniques. As 
discussed by \cite{Greene2005}, the NLR is sufficiently compact to be
illuminated by the AGN, while large enough to feel the gravitational
potential of the bulge. Therefore, there is a relatively good
correlation between the gas and the stellar velocity dispersion around
local AGN. In the mid-IR \cite{Dasyra2008,Dasyra2011} showed that the
velocity dispersion of the fine structure lines \Siv, \Neiii, \Neva,
and \Oiv \, of AGN are well correlated with $M_{\rm BH}$.

For our {\it
  Spitzer}/IRS spectra we focus on the \Neiii
\, line to look for spectrally resolved line profiles. In AGN the
flux and luminosity of this
fine-structure line are found to correlate well with those of the mid-IR
[Ne\,{\sc 
  v}] lines at 14.3 and $24.3\,\mu$m \citep{Gorjian2007, 
Pereira2010lines}. The high ionization potential of
the [Ne\,{\sc 
  v}] lines indicates that they are mostly excited by  AGN and thus, 
the \Neiii \, emission is likely to be as well. When
compared with 
other mid-IR fine structure lines with relatively high ionization
potentials (e.g., \Siv, \Oiv, and the [Ne\,{\sc
  v}] lines), the \Neiii \, line is a
good compromise between ionization potential, critical density\footnote{For
  reference the \Neiii \, line has a slightly higher ionization
  potential than the optical [O\,{\sc iii}]$\lambda$5007 emission line
but the optical line has a slightly higher critical density \citep[see figure~6 of
][]{Dasyra2011}.},
intensity of the line, and a clean region of the mid-IR spectrum for
accurate measurements.

Using the calibration sources (Section~\ref{sec:calsources}) we obtained twelve independent
measurements of the unresolved width of the \Neiii \, line and 
inferred an instrumental width of  
${\rm FWHM}_{\rm inst} = 0.02618\pm0.0007\,\mu$m. This is equivalent to a 
spectral resolution of $R=\lambda/{\rm FWHM}_{\rm inst} \sim 595 \pm
16$ and an instrumental velocity dispersion of $\sigma_{\rm inst} =
215 \pm 6\,{\rm km\,s}^{-1}$ at the wavelength of this line.
 While the derived spectral resolution is very
similar to that from \cite{Dasyra2011}, the standard deviation of our
measurements is much lower. \cite{Dasyra2011} obtained $\sigma_{\rm
  inst}$ as the average of the values of the high excitation fine 
structure lines of a sample of AGN that were deemed not to be
resolved by {\it Spitzer}/IRS with their method. By doing so, they 
probably also 
included measurements of barely resolved lines. This would therefore 
explain the higher dispersion of their  instrumental value.

For the LIRGs we  need a criterion to assess whether 
the lines are resolved or not since a significant number of sources
have measured widths slightly larger than the instrumental
resolution. Figure~\ref{fig:FWHMdistribution}
shows the distribution of 
the measured FWHM of the \Neiii \, line for the sample of LIRGs
compared with those of the calibration sources.  
To avoid analyzing
barely resolved lines we adopt the following criterion for the
velocity dispersion to determine
whether a line is clearly resolved: $(\sigma_{\rm obs} - \sigma_{\rm inst}) > 3\times\sqrt{\epsilon_{\rm m}^2 +
\epsilon_{\rm inst}^2}$, where $\sigma_{\rm inst}$ and
$\epsilon_{\rm inst}$ are the instrumental resolution and its standard
deviation, and $\sigma_{\rm obs}$ and $\epsilon_{\rm m}$ are the measured
values for the LIRGs. The total error of the observations $\epsilon_{\rm obs}$ includes
both the error in the measurement $\epsilon_{\rm m}$ and that associated with the
instrumental value, that is, $\epsilon_{\rm obs} =(\epsilon_{\rm m}^2 +
\epsilon_{\rm inst}^2)^{1/2}$. 

\begin{table*}
\small
\center
\caption{Velocity dispersions and BH masses }\label{table:sigmas}
\begin{tabular}{lcccccccc}
\hline
\hline
Galaxy          & Class  & $\sigma$([NeIII]) & 
$\sigma$([OIII])$_{\rm n}$ & $\sigma$([OIII])$_{\rm b}$ & $\sigma_{\rm *}$ & Ref & $\log {\rm M}_{\rm BH}$ &\\
 & & km s$^{-1}$ &  km s$^{-1}$ & km s$^{-1}$ &km s$^{-1}$  & & M$_\odot$ \\
\hline
\multicolumn{8}{c}{Seyfert nuclei and [NeV] emitters}\\
\hline
CGCG~468-002-NED01 & Sy2$^*$ & $135\pm 12$ & $218\pm22$ & \nodata & \nodata&
\nodata & 8.06\\
NGC~3690         & Sy2 & $167\pm10$  & \nodata  & \nodata & $144\pm11$
& 1 & 7.52\\
MCG~$-$03-34-064   & Sy1 & $368\pm18$  & $249\pm10$ & $656\pm26$ & 155 
& 2 & 7.65\\
NGC~5135         & Sy2 & $138\pm9$  & $127\pm9$    & $305\pm21$ & $124\pm6$ & 
3 & 7.24 \\
IC~4518W         & Sy2 & $156\pm12$   &\nodata & \nodata & \nodata &
\nodata &7.48 & \\
NGC~5990         & Sy2 & $264\pm11$ &\nodata & \nodata & \nodata &
\nodata & $<8.45$ &   \\
NGC~6156         & [Ne\,{\sc v}]& $160\pm13$  &\nodata & \nodata & \nodata &
\nodata & 7.52 \\
MCG~+04-48-002   & [Ne\,{\sc v}] & $158\pm7$  &\nodata & \nodata & \nodata &
\nodata & 7.50   \\
NGC~7130         & Sy2: &  $259\pm16$  & $130\pm9$ & $463\pm32$ & $147\pm5$ & 
3  & 7.55\\
NGC~7469         & Sy1  & $182\pm13$  & $145\pm7$ & $336\pm17$ & $152\pm16$ & 
4 & 7.61\\
NGC~7591         & Sy2: & $169\pm17$  &\nodata & \nodata & \nodata &
\nodata & 7.62  \\
NGC~7679 & Sy1/Sy2 & $102\pm7$ & 176$^*$ & \nodata & 96 & 
2 & 6.77\\
\hline
\multicolumn{8}{c}{Composite and H\,{\sc ii} nuclei}\\
\hline
NGC~1614 & Composite & $157 \pm 10$ & \nodata  & \nodata  & $164\pm8$
& 5  & 7.75\\ 
NGC~2369         & Composite & $133 \pm 8$ & $146\pm15$ & \nodata & \nodata & \nodata &
\nodata &   \\
IC~694 & LINER & $<70$ & \nodata & \nodata  & $141\pm 17$ & 5 & 7.48\\
NGC~3256         & H\,{\sc ii} & $<112$ & \nodata & \nodata &
$100\pm6$ & 
3 & 6.84\\
MCG~$-02$-33-098W & Composite & $103 \pm 9$ & \nodata & \nodata & \nodata &
\nodata &  \nodata \\
ESO~173$-$G015 & \nodata & $102\pm7$ & \nodata & \nodata & \nodata &
\nodata &  \nodata \\
IC~4280         & H\,{\sc ii}  & $128\pm 8$ & \nodata & \nodata & \nodata &
\nodata &  \nodata \\
IRAS~17138$-$1017N & Composite & $<80$ &\nodata &\nodata  &
$72\pm 5$ & 6 & 6.24 \\ 
IRAS~17578$-$0400 & \nodata & $119\pm 8$ & \nodata & \nodata & \nodata &
\nodata &  \nodata \\
NGC~6701         & Composite & $<159$ & \nodata & \nodata &
$150\pm 20$ & 7 & 7.59 \\ 
NGC~7771         & Composite & $178\pm17$  &\nodata & \nodata & \nodata &
\nodata &  \nodata \\
\hline
\end{tabular}

{\bf Notes.}--- The references for the spectral class of the nuclei are
listed in
\cite{AAH12} except for CGCG~468-002-NED01,
which is from the newly analyzed
optical spectrum (see Section~\ref{sec:oiiiline}).\\
$^*$The [O\,{\sc iii}]
velocity dispersion for NGC~7679 is also from reference 2.\\
References for $\sigma_{\rm *}$: 
1. \cite{Ho2009}. 
2. \cite{Gu2006}. 
3. \cite{GarciaRissmann2005}. 
4. \cite{Onken2004}. 
5. \cite{Hinz2006}.  
6. \cite{Shier1998}.
7. \cite{Marquez1996}. 
\end{table*}

Using the criterion described above we
found that 19 LIRGs in our sample show  resolved \Neiii \,
lines. Most  are 
LIRGs classified as Seyfert (Figure~\ref{fig:FWHMdistribution}) and/or
are [Ne\,{\sc v}] emitters. Those nuclei classified as H\,{\sc ii}, on
the other hand, tend to show FWHM close to
the instrumental resolution. Composite nuclei have FWHM  between those
of unresolved lines and the clearly resolved \Neiii \, lines. For
  those LIRGs deemed to have unresolved \Neiii \, lines, the average
  measured FWMH of the \Neiii \, line is
  $0.02776\pm0.0017\,\mu$m. That is,  the observed values for these
  lines are on
  average 2$\sigma$ above the instrumental resolution.

Table~\ref{table:sigmas} lists the intrinsic values of the velocity
dispersion for the LIRGs with resolved \Neiii \, lines as well as
their spectroscopic classes from optical spectroscopy and/or mid-IR
indicators \citep{AAH12}. The velocity dispersions (corrected for
instrumental resolution) are between 102 and $368\,{\rm
  km\,s}^{-1}$. For galaxies in our sample with literature
values of the stellar velocity dispersions whose \Neiii \, lines are
deemed to be unresolved, we provide in this table upper limits to the
gas velocity 
dispersion. 
In  Figure~\ref{fig:profiles} we show the normalized profiles of the twelve
Sy/[Ne\,{\sc v}] emitter LIRGs with spectrally resolved \Neiii \,
lines compared with those of 
an unresolved 
profile represented as a Gaussian function.

\subsection{Optical [O\,{\sc iii}]$\lambda$5007 measurements}\label{sec:oiiiline}

We obtained  archival optical spectra of six LIRGs classified as AGN or
Composite from the
six-degree Field (6dF) Galaxy Survey  
 \citep[6dfGS,][]{Jones2004, Jones2009}
and from observations with the FAST spectrograph
\citep{Tokarz1997, Fabricant1998}  
from the  Astrophysics (CfA) telescope data
center. 

NGC~2369, MCG~$-$03-34-06, NGC~5135, and NGC~7130 were observed 
with the 6dF multi-object fibre (angular diameter of 6\farcs7)
spectrograph on the United Kingdom Schmidt Telescope (UKST). At the typical 
distance of these LIRGs the spectra cover the central $\sim$2\,kpc. The 
instrument spectral
resolution around the [O\,{\sc iii}] emission lines is
$\sigma_{\rm inst} \sim$125\,km\,s$^{-1}$. 
 Long-slit spectroscopy  was
available for CGCG~468-002-NED01 and NGC~7469 taken with  FAST  on the
Mt. Hopkins Tillinghast 60-inch telescope as part of different
observing programs. The slit width was 3\arcsec\ and the spectral
resolution was $\sigma_{\rm inst} \sim$75\,km\,s$^{-1}$. None of the
spectra are flux 
calibrated, but they can be used to measure the velocity dispersion 
of the emission lines as well as their ratios. 

We first measured the optical line ratios of CGCG~468-002-NED01 since
we did not have an optical 
classification for this galaxy. We classify this galaxy as a
Seyfert 2.  In \cite{Pereira2010lines} and 
\cite{AAH12} we detected the mid-IR [Ne\,{\sc v}] lines in  this galaxy
  and measured a high  \Oiv/\Neii \, line ratio. We used both as
  evidence of the presence of an AGN in this 
nucleus. The optical classification confirms this.

For all the galaxies we modeled the [O\,{\sc iii}] emission lines at
5007\AA\ and 4959\AA\  by 
fitting a Gaussian profile to each line. The relative position of the
Gaussians was fixed according to the rest-frame wavelengths of the two
lines. Likewise, the relative intensity was also fixed to the value
determined by the atomic parameters.
We found that the fit with
one component was only satisfactory for CGCG~468-002-NED01 and
NGC~2369. For the rest we had to add a broad 
component to the [O\,{\sc iii}] lines. The wavelengths and
intensities of the narrow (also called core) component and the 
broad component were considered
independent. To estimate the errors of the velocity dispersions
we used the rms of the continuum adjacent to the line for the 6dF
spectra and the errors of the spectra for the FAST data. 
The typical errors of the velocity
dispersions of the [O\,{\sc iii}] lines are less than 10\%, and are listed for 
each galaxy in Table~ \ref{table:sigmas}.

The top panel of  Figure~\ref{fig:profilesOIII}
shows the galaxies for which the [O\,{\sc iii}]$\lambda$5007 line fits
required two components,  whereas the bottom panel shows the
  profiles of NGC~2369 and CGCG~468-002-NED01. Table~\ref{table:sigmas} 
lists the velocity dispersion of
the core of the line and the broad component, if present, corrected
for instrumental resolution. The
velocity dispersions of the core components measured in the four
galaxies with double components are in good agreement with
literature $\sigma_{\rm *}$ values,  except for MCG~$-$03-34-064.

For the four LIRGs with double components 
the broad component is blueshifted typically by 
$50-300\,$km\,s$^{-1}$. The presence of such 
asymmetric or broad blue wings in the  
[O\,{\sc iii}]$\lambda$5007 lines is common in AGN \citep[see
e.g.,][]{Greene2005}. 
The total [O\,{\sc iii}] emission is dominated by
the broad component in MCG-03-34-064 and in NGC~7130, whereas in 
NGC~5135 and NGC~7469 the core component dominates the
emission. 
In NGC~5135 \cite{Bedregal2009} based on both the 
relatively broad [Si\,{\sc
vi}]$1.96\,\mu$m coronal line and its spatial extent suggested the
existence of different kinematical components and the possible
presence of AGN induced outflows. \cite{GonzalezDelgado1998} found
blueshifted components in the 
Ly$\alpha$ and [O\,{\sc iii}]$\lambda$5007 lines of both NGC~7130 and
NGC~5135. The velocity dispersion of the broad component of 
the [O\,{\sc iii}]$\lambda$5007
line of NGC~7130 is similar to the H$\alpha$ broad component  
\citep[$\sigma \sim 400\,{\rm
  km\,s}^{-1}$, ][]{Bellocchi2012}.  
In NGC~7469 \cite{Wilson1986} also detected two components in the 
[O\,{\sc iii}]$\lambda$5007 line.

\vspace{0.5cm}

\section{Results}\label{s:results}

\subsection{Black Hole Masses}\label{s:resolvedlines}
In Table~\ref{table:sigmas} we list,
in addition to the gas velocity dispersions from 
the \Neiii \, and optical [O\,{\sc iii}]$\lambda$5007 lines, 
the  stellar velocity dispersion $\sigma_{\rm *}$ values from the
literature, when
available. We also included in this table  $\sigma_{\rm *}$ values
from the literature for
other LIRGs in our sample (IC~694, NGC~3256, IRAS~17138$-$1017, 
and NGC~6701) whose \Neiii 
\, lines appear unresolved at the SH spectral resolution. We note,
however, that in the case of on-going major mergers the stellar
velocity dispersion vs. BH mass relation might not be applicable.

\cite{Dasyra2008, Dasyra2011} showed that mid-IR resolved fine-structure
lines can be used to study the NLR of AGN. In
particular they  demonstrated that the mass of the BH correlates with
both the velocity dispersion 
and the luminosity of the NLR region for a sample of optically
selected AGN. The relation is: 

\begin{equation}
\log{\rm (M_{\rm BH}/{\rm M}_\odot}) = -1.82 + 4.24\,\log (\sigma_{\rm
NeIII}/{\rm
  km\,s}^{-1}) 
\end{equation}

\cite{Dasyra2011} fitted a 
similar relation for the optical [O\,{\sc iii}]$\lambda$5007 line:

\begin{equation}
\log{\rm (M_{\rm BH}/{\rm M}_\odot}) = -1.78 + 4.24\,\log (\sigma_{\rm
OIII}/{\rm
  km\,s}^{-1}) 
\end{equation}

For both relations  \cite{Dasyra2011} fixed the slope to the
value derived from the correlation between  the 
stellar velocity dispersion and the BH mass by \cite{Gultekin2009}:

\begin{equation}
\log{\rm (M_{\rm BH}/{\rm M}_\odot}) = 8.12 + 4.24\,\log (\sigma_{\rm
*}/200{\rm
  km\,s}^{-1}) 
\end{equation}

\noindent The typical uncertainties on the derived $M_{\rm BH}$ can be as
high as 0.8\,dex if using  the gas 
$\sigma$ based on
the rms of the relations \citep{Dasyra2011}
and 0.4\,dex if using 
$\sigma_{\rm *}$ \citep{Gultekin2009}.

Before we compute the BH masses, we compare the gas velocity
dispersions for the LIRGs with both \Neiii \, and [O\,{\sc
  iii}]$\lambda$5007 measurements. As we saw in Section~\ref{sec:oiiiline},
in some cases the optical line is fitted with two components.  For AGN
there is generally a good agreement between $\sigma_{\rm *}$ and the gas velocity
dispersion from the narrow component of [O\,{\sc iii}] 
\cite[][]{Onken2004, Greene2005}. Therefore,  
the core of the [O\,{\sc iii}]$\lambda$5007 line probes
the gravitational movements in the NLR and thus can be used to
estimate $M_{\rm BH}$.  On the other hand, the broad components  
tend to be blueshifted, may be more influenced by the AGN, 
and are likely produced by 
outflows.  As can be seen from Table~\ref{table:sigmas},
the velocity dispersions from the \Neiii \, line appear to be intermediate
between those of the two optical components. Spectrally resolved
blueshifted \Neiii \, lines  
(as well as the [Ne\,{\sc v}] lines) have also been reported for local ULIRGs
\citep{Spoon2009}, and have been interpreted as the
result of outflows.  Thus, it is possible that
in some cases we might overestimate the BH masses when
using the \Neiii \, line if there are several components. The most
suspect galaxy in our sample would be NGC~5990, for which $\sigma_{\rm
NeIII}$ predicts the most massive BH by far in our sample. 
We note, however, that the
calibration of Dasyra et al. (2011) should take these effects into
consideration.

In  
Table~\ref{table:sigmas} we list $M_{\rm BH}$ as
calculated from, in this order of preference, the 
stellar velocity dispersion, the velocity
dispersion of the core of the [O\,{\sc iii}]  line, 
and the \Neiii \, line velocity dispersion. 
For those LIRGs with no definitive
evidence of AGN activity (i.e., classified as 
composites, H\,{\sc ii} or unknown), we only
list $M_{\rm BH}$ in the second part of the table 
if we have a value of $\sigma_{\rm *}$. This is because we cannot
be sure whether the \Neiii \, emission comes mostly from the NLR or
whether it is produced by star formation \citep{Pereira2010lines}.

For those LIRGs hosting an
actively accreting BH (that is, with clear signs of 
AGN activity, first part of Table~\ref{table:sigmas}) we find typical
BH masses of $3\times 
10^{7}\,M_\odot$, ranging from $6 \times 10^{6}\,M_\odot$
to  $3\times 10^{8}\,M_\odot$. The estimated BH masses for those LIRGs
classified as composite or H\,{\sc ii} are also in this range.
The masses of the LIRG BHs  appear to be 
similar to those of the currently
growing BH in the local universe  hosted in late-type 
galaxies \citep{Heckman2004, Schawinski2010}.

The typical Eddington ratios
($L_{\rm bol}({\rm 
  AGN})/L_{\rm Edd}$)  for the LIRGs
with an estimate of $M_{\rm BH}$ are $2 \times 10^{-2}$, and range  between
$5\times 10^{-4}$ and 0.1. This means that the BH in local LIRGs are
accreting at a lower efficiency than those in local ULIRG
\citep[typical values of $0.08-0.4$ depending on the method used for
estimating the BH mass, ][]{Dasyra2006,Veilleux2009}.

The majority of LIRGs with spectroscopically resolved 
\Neiii \, lines are classified as Seyfert galaxies. However, in local
LIRGs there is a high fraction of composite objects
(AGN/starburst). Moreover, this fraction remains approximately constant as
a function of IR luminosity from galaxies with $L_{\rm IR} \sim
10^{10}\,L_\odot$ up to ULIRGs
\citep{Yuan2010}. As can be seen from Figure~\ref{fig:FWHMdistribution}, 
composite nuclei in local LIRGs have values of the FWHM of \Neiii \,
on average lower than those of Seyfert nuclei but higher than those of
nuclei classified as H\,{\sc ii} and the instrumental resolution. 
Assuming that these composite nuclei do indeed host an AGN and that
the \Neiii \, line is probing their NLR, then their
BH masses ought to be less than $\sim 2-3\times 10^{7}\,M_\odot$. 
The estimated typical AGN bolometric luminosities from the mid-IR
spectral decomposition are $10^{43}\,{\rm erg \, s}^{-1}$ or less, so
we cannot constrain their Eddington ratios.

The BH of local LIRGs are  only marginally less massive than those of local
ULIRGs. The BH masses of the latter are in the range 
$\sim 10^{7}-5\times 10^{8}\,M_\odot$ \citep{Tacconi2002, Dasyra2006,
  Veilleux2009}. Samples of local ULIRGs are mostly dominated
by gas-rich  interacting galaxies and major mergers
\citep{Sanders1996}, with coalesced ULIRGs having slightly larger BH
masses than pre-coalescence ULIRGs \citep{Dasyra2006}. Our sample is
both flux and volume-limited and thus is composed mostly of LIRGs 
with 
 $L_{\rm IR} \sim 1-2 \times 10^{11}\,L_\odot$ (individual galaxies), 
which also tend to be spiral galaxies, minor mergers, and galaxies in
groups. That is, morphologically local LIRGs are not dominated in
numbers by
major mergers \citep[see e.g.][]{Sanders2004, Kaviraj2009, 
PereiraSantaella2012} and their BH masses are similar to those of
pre-coalescent ULIRGs \citep{Dasyra2006}.  
%

\begin{figure*}
\hspace{2cm}
\includegraphics[width=0.38\textwidth,angle=-90]{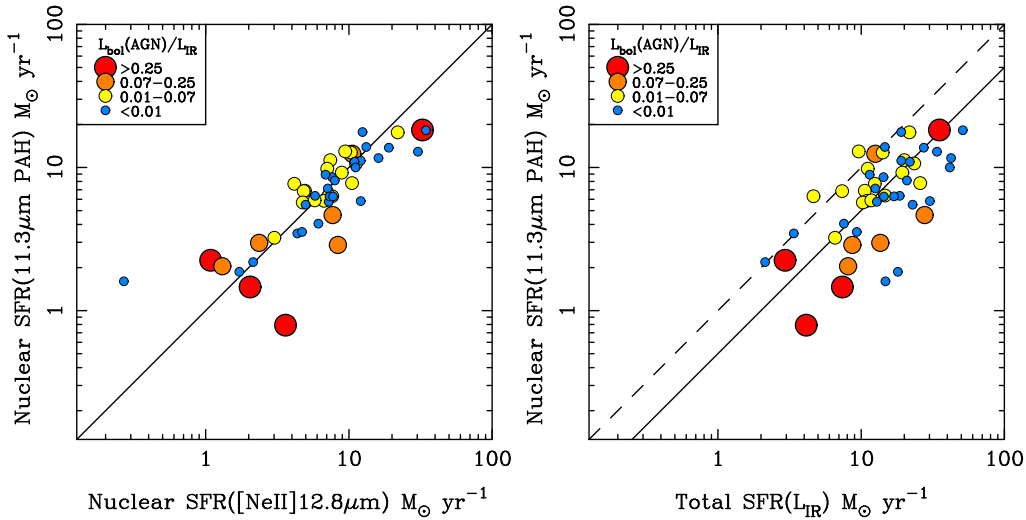}

\caption{Left panel: comparison between the nuclear SFRs
  from the \Neii \, line (corrected for AGN emission)
and the $11.3\,\mu$m PAH feature
  for the sample of local LIRGs. The colors denote the 
AGN bolometric contribution to $L_{\rm IR}$ of the individual
  galaxies: 
$<1\%$ and no AGN contribution (blue),  $1-7\%$ (yellow),
$7-25\%$ (orange), and $>25\%$ (red). 
The highly discrepant point with no AGN contribution to the left of
the straight line is IC~860. 
The straight solid line is a 1:1 line and is the average value of the
ratio between 
the nuclear SFRs computed with the two estimators (see the text).
Right
                         panel: comparison between the nuclear SFR from
                       the $11.3\,\mu$m PAH feature and the integrated
                       SFR from the IR luminosity after subtracting
                       the AGN contribution as estimated by
                       \cite{AAH12}. The straight solid line shows an average
                       ratio between the nuclear and the total SFRs of 0.5
                       (see text), typical of local LIRGs, whereas the
                       dashed line represents a 1:1 relation.
}
\label{fig:SFRs}
\end{figure*}

\subsection{Star Formation Rates}\label{sec:SFR}
We computed the SFRs for our sample of LIRGs using a number of IR-based 
indicators, including the fine-structure \Neii \, line
\citep{Roche1991, HoKeto2007},  the $11.3\,\mu$m  PAH
feature \citep{Brandl2006,DiamondStanic2012}, and the total
$8-1000\,\mu$m IR luminosity 
\citep{Kennicutt1998}. Although it is not clear whether
the presence of an AGN could destroy the PAH carriers or not, 
at the typical luminosities of the AGN hosted in local LIRGs
the
$11.3\,\mu$m PAHs do not seem to be affected
\citep{DiamondStanic2010}. 

For the integrated SFRs we used $L_{\rm IR}$ of the individual galaxies as given in \citet[][their
table~1]{AAH12} after subtracting the AGN
contribution. We used the
\cite{Kennicutt1998} relation in terms of the IR luminosity 
converted to a Kroupa IMF:

\begin{equation}\label{eq:SFRIR}  
{\rm SFR} (M_\odot \;{\rm yr}^{-1})=1.14\times
10^{-10}\;(L_{\rm IR}, L_\odot).
\end{equation}

\noindent We obtained total SFRs for the individual galaxies in our sample of
between 1 and $60\,M_\odot\,{\rm yr}^{-1}$. 

For the nuclear SFR we used the {\it Spitzer}/IRS spectroscopy. 
At the median distance of 65\,Mpc for
our sample of LIRGs, the IRS SH slit width (4.7\,arcsec) 
subtends typically 1.5\,kpc. We  use
the following recipes put forward by \cite{DiamondStanic2012} to
compute the nuclear  
SFRs based on the \Neii \, line $11.3\,\mu$m PAH feature luminosities:

\begin{equation}\label{eq:SFRPAH} 
{\rm SFR}(M_\odot \;{\rm yr}^{-1})=9.6\times
10^{-9}\;L(11.3\,\mu{\rm m}, L_\odot) 
\end{equation}

\begin{equation}\label{eq:SFRNeII} 
{\rm SFR}(M_\odot \;{\rm yr}^{-1})=8.9\times
10^{-8}\;L({\rm [NeII]}, L_\odot). 
\end{equation}

\cite{DiamondStanic2012} calibrated Equations~\ref{eq:SFRPAH} and
\ref{eq:SFRNeII} for galaxies with  $L_{\rm IR}< 10^{11}\,L_\odot$ using the
\cite{Rieke2009} templates and 
a Kroupa IMF. We chose to use these calibrations rather than those for
higher IR luminosities because the median
value of the IR luminosity of the individual galaxies in our sample is 
$1.3\times 10^{11}\,L_\odot$. In addition, the width of the 
IRS SH slit only probes the
nuclear regions of our sample of LIRGs 
and thus lower IR luminosities.

We also note that Equation~\ref{eq:SFRPAH}
assumed that the flux of the 11.3 $\mu$m PAH feature was measured using 
{\sc pahfit} \citep{Smith2007}. The spectral coverage ($\sim 10-18\,\mu$m)
and spectral resolution of the SH  
data used here are not adequate to use {\sc pahfit}. We measured the
PAH fluxes using a local continuum
(Section~\ref{s:observations} and
  Table~\ref{table:PAH}). Following \cite{Smith2007} we applied 
a multiplicative 
factor of two to the PAH fluxes measured with a local continuum 
to make a proper comparison
with {\sc pahfit} fluxes. We also corrected the
$11.3\,\mu$m PAH fluxes for extinction using the nuclear strength  of the
$10\,\mu$m silicate feature \cite[from][]{AAH12} and the extinction law given in
\cite{Smith2007}. For most of our local LIRGs
 this correction is small, as the silicate
absorptions are moderate \citep[see e.g. figure~6 in][]{AAH12} as
compared with the more deeply embedded population of local ULIRGs
\citep{Spoon2007}.

While the \Neii \, emission in  
low luminosity and moderate luminosity AGN is mostly produced by star
formation, in high luminosity AGN this emission line 
can also have an important
contribution from the AGN 
\citep{Pereira2010lines}. To correct  the \Neii \, line flux for
  any possible AGN contribution,  
  we followed the method put forward by
  \cite{Veilleux2009}. This method reproduces the observed values of
  the \Oiv/\Neii 
  \, line ratio \citep[given in ][]{AAH12} as a fractional 
combination of the typical line ratio
  of an AGN ($\sim 4$) and that of a starburst galaxy ($\sim 0.01$),
  where the AGN contribution is the free parameter.
Using this method we find that the AGN contribution to
  the nuclear  \Neii \, emission of local LIRGs is small. These
  corrections are only 
necessary for 10 LIRGs and the AGN contribution to 
the nuclear \Neii \, fluxes ranges from 3\% to
53\% (average $\sim 18\%$).

 Figure~\ref{fig:SFRs} (left panel) compares the
nuclear SFRs as derived from the \Neii \, line  (after removing
  the AGN contribution) and from the $11.3\,\mu$m PAH
feature. We have  color coded symbols by the AGN
bolometric contribution to IR luminosity of the galaxies
\citep[see][]{AAH12}.  As can be seen from this
figure, the majority of LIRGs with small or no ($<$25\%) AGN contribution 
show a good agreement between the derived nuclear
SFRs using the two indicators. Indeed, we find that the ratio 
between the nuclear SFR estimated from the $11.3\,\mu$m PAH feature
and that from the \Neii \, line is on average $1.0\pm 0.3$.  

In a few cases with a significant AGN  bolometric contribution
the nuclear SFR derived 
from the \Neii \, line is higher than that from the $11.3\,\mu$m PAH
feature, even after correcting for the AGN contribution.
Given the good agreement between the
nuclear SFR computed with the two methods  for LIRGs with small or
  no AGN contribution and to avoid AGN
contamination issues from now on for the nuclear SFR we use those
derived from the $11.3\,\mu$m PAH feature. 
Thus, the individual nuclei of the sample of local  
LIRGs show values of the nuclear SFR of between $\sim 0.8$ and $\sim 20\,
  M_\odot \,{\rm yr}^{-1}$.

In  Figure~\ref{fig:SFRs} (right) we compare the total SFR
against the nuclear (typically on scales of 1.5\,kpc) SFR for the
individual galaxies of the LIRG sample.  It is clear that 
a significant fraction of the individual galaxies in our
sample have more than half of their total SFRs taking place 
outside the nuclear ($1-2\,$kpc) regions. For the
individual galaxies of the  
sample the average value of the
nuclear SFR over the total SFR ratio is $0.5\pm 0.3$. This agrees with
findings using other indicators 
\citep{Hattori2004, AAH06, DiazSantos2010, RodriguezZaurin2011}. We
also note, however, that in 
approximately 20\% of our volume-limited sample of LIRGs 
the nuclear emission accounts
for most of the total SFR measured in the individual galaxies, as is
the case for a large fraction of local ULIRGs.

Incidentally, the same comparison for the nuclear SFRs 
(Figure~\ref{fig:SFRs}, left) but using the SFR equations calibrated for 
$L_{\rm IR} > 10^{11}\,L_\odot$ \citep[see][]{DiamondStanic2012}, 
results in  nuclear SFRs  from the $11.3\,\mu$m
PAH feature which are on average a factor of 1.3 higher than those
 from the \Neii \, line. Moreover, the ratio between the 
nuclear SFR from the $11.3\,\mu$m
PAH feature and the total SFR from $L_{\rm IR}$ would also be 
on average 1.3. This gives us confidence that we used the appropriate
SFR calibrations for the nuclear rates.

\subsection{Relation between BH accretion rates and SFR in LIRGs}\label{sec:BHAR}
The presence of an AGN is an unambiguous signpost of a period of BH
growth. The AGN luminosity can then be expressed in terms of the 
BHAR $\dot m_{\rm
  BH}$ and the mass-energy conversion efficiency $\epsilon$
\citep[see][and references therein]{Alexander2012}: 

\begin{equation}\label{eq:BHAR}
\dot m_{\rm BH}({\rm M}_\odot\,{\rm yr}^{-1})=0.15(0.1/\epsilon)(L_{\rm bol}/10^{45}\,{\rm erg\,s}^{-1})
\end{equation}

We used this relation with the typical value of $\epsilon=0.1$ in the
local universe \citep{Marconi2004} and  $L_{\rm bol}({\rm AGN})$ from
the mid-IR spectral decomposition \cite[see][]{AAH12}.  
We obtained BHAR  between 0.0007 and
$0.08\,M_\odot\,{\rm yr}^{-1}$ for local LIRGs. The uncertainties
  of the estimates of the BHAR are dominated by the uncertainties of
  the AGN bolometric luminosities which are typically less than 0.4dex
\citep[see][for details]{AAH12}.

Having calculated the BHAR we can now compare them with the nuclear
and total SFR for the sample of LIRGs. The distribution of  
SFR/$\dot m_{\rm BH}$  can provide clues as to whether AGN and on-going star
formation activity are contemporaneous in local LIRGs. Recent numerical
simulations by \cite{Hopkins2012} predict a  time 
offset between the peaks of these activities which is  thought to
depend on the physical scale within the galaxy and the dynamical time
of the galaxy, among other parameters. 

\begin{figure}
\includegraphics[width=0.5\textwidth,angle=-90]{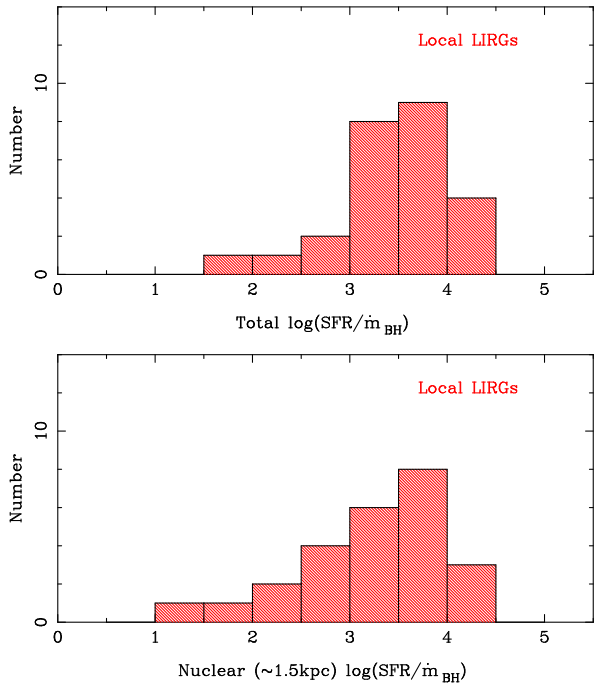}

\caption{Distributions of the SFR/$\dot m_{\rm BH}$  ratios
for  nuclear (typical scales of 1.5\,kpc) SFRs  (lower panel) 
and integrated SFRs 
(upper panel) for the sample of local LIRGs with an estimate of the AGN
  luminosity. The nuclear SFR are from the $11.3\,\mu$m PAH
  feature and the integrated SFR from $L_{\rm IR}$ after
  subtracting the AGN component (see the text for details). }
\label{fig:SFRoverBHAR}
\end{figure}

Figure~\ref{fig:SFRoverBHAR}
shows the distribution of  SFR/$\dot m_{\rm BH}$ for the sample of
local LIRGs for the nuclear ($\sim 1.5\,$kpc) and the integrated
SFRs, only for those LIRGs with an estimate of 
$L_{\rm bol}({\rm AGN})$. The nuclear and integrated SFR/$\dot m_{\rm BH}$ 
are in the range $\sim 12$ to $\sim 2\times 10^{4}$,
with median values of the nuclear and integrated $\log ({\rm SFR}/\dot
  m_{\rm BH})$ of 3.1
and 3.4, respectively. Most of the nuclei classified as Seyfert from
optical spectroscopy 
show nuclear  $\log ({\rm SFR}/\dot
  m_{\rm BH})<3.0$. The nuclei classified
as composite tend to show larger values of the nuclear SFR to BHAR
ratio, as expected if they were powered by both SB and AGN
activity.

For the entire sample of local LIRGs we obtain
$\sum {\rm SFR}/\sum \dot m_{\rm BH} \geq 2900$. This is a lower
limit because for half of the sample we did not detect an AGN
component and therefore we are only using an upper limit for their
BHAR. If we repeat this  only for 
those LIRGs with an estimate of their AGN bolometric
luminosities we obtain a time-averaged value of 
${\rm SFR}/\dot m_{\rm BH} \simeq 2500$, if we assume
that all LIRGs go through an AGN phase, that is, a $\sim 50\%$ duty
cycle. The two values are consistent with each other taking into
account the typical uncertainties in calculating $L_{\rm bol}(AGN)$
\citep[$\sim 0.4\,$dex,][]{AAH12}. 
This ratio is a few times higher than that measured 
for the
local population of bulge-dominated galaxies  
\citep[$\sim 10^3$, ][]{Heckman2004} and the 
local normalization of the BH mass versus bulge mass relation
\citep{Marconi2003, Haring2004}.
Even though the SFRs of LIRGs are much higher than those of local
bulge galaxies, the time-averaged ${\rm
  SFR}/\dot m_{\rm BH}$ value of local LIRGs is  similar to that
of local bulge galaxies during the first 0.3\,Gyr of their on-going starburst
\citep[][their figure~9]{Wild2010}. This suggests the existence of similar time delays
between the peak of star formation and BH growth in local LIRGs as well.

\section{Discussion}

\subsection{Comparison between the AGNs in LIRGs and  the optically
  identified RSA Seyferts}\label{sec:comparison}  
In this section we 
compare the properties of the AGNs identified in
local LIRGs with the optically selected 
Seyfert galaxies in the revised-Shapley-Ames
catalog \citep[RSA,][]{Sandage1987,Maiolino1995}. 
This optical sample was
spectroscopically selected from the original galaxy-magnitude limited
sample of RSA galaxies and is believed not to be biased against
low-luminosity Seyfert galaxies. Also, as for the sample of local
LIRGs studied in this work, the Seyfert activity in the RSA sample
does not appear to be 
driven primarily by mergers for the majority of the galaxies. Finally, the 
AGN bolometric luminosities of the RSA Seyferts are similar to those
of the local LIRGs. 


\begin{figure}
\includegraphics[width=0.5\textwidth,angle=-90]{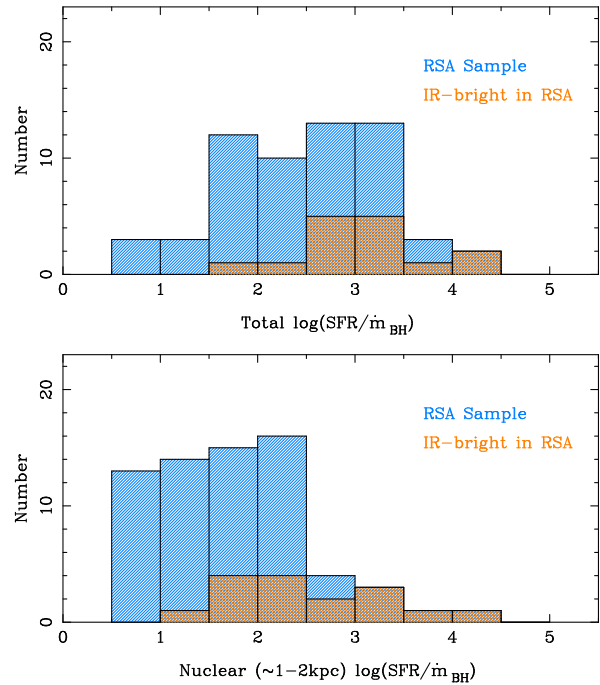}
\vspace{0.25cm}

\caption{Distributions of the ratio between the nuclear SFR and the BHAR 
  (bottom panel) and the integrated SFR and the BHAR
  (upper panel) for the RSA sample of AGN (blue). In orange we mark
  those RSA AGN that are also IR-bright galaxies (see text). We 
  only included RSA galaxies 
  with $\dot m_{\rm BH} > 10^{-4}\,M_\odot\,{\rm yr}^{-1}$.}
\label{fig:SFRoverBHAR_onlyRSA}
\end{figure}
\smallskip

The median values of ${\rm
  SFR}/\dot m_{\rm BH}$ for our LIRGs are on average
higher than those of local Seyfert galaxies, especially the nuclear
values \citep[see
Figure~5 and][]{DiamondStanic2012}. Of course, this is not completely
unexpected as local LIRGs, including those hosting an AGN, 
are essentially selected by their star formation activity, whereas the
RSA Seyferts are selected by their host galaxy magnitudes. The latter
is more likely probing the stellar mass, rather than the star
formation activity.

We can make a  more meaningful comparison if we identify the IR-bright
Seyferts in the RSA sample. We set the limit
to $\log ({\rm L}_{\rm IR}/L_\odot) = 10.8$ to match approximately the lower
limit covered by 
the IR luminosities of the individual galaxies of our sample of local
LIRGs \citep[see table~1 in][]{AAH12}. We also included from our
sample the newly identified Seyfert nuclei with  an estimate of $L_{\rm bol}({\rm AGN})$.
The  values of SFR/$\dot m_{\rm BH}$ for the RSA Seyferts are taken from 
\cite{DiamondStanic2012} except for those already included in our sample of
LIRGs. 
\cite{DiamondStanic2012} computed the nuclear SFRs 
from the $11.3\,\mu$m PAH feature and the MIPS $24\,\mu$m photometry, 
and $L_{\rm bol}({\rm AGN})$ from the \Oiv \, line. Finally,
 for the  few RSA Seyferts
deemed to have their \Oiv \, luminosities strongly contaminated by
star formation, we estimated  $L_{\rm bol}({\rm AGN})$ (and thus
BHAR) from their hard X-ray luminosities and using a bolometric correction. 
In this comparison we only included RSA Seyferts with $\dot
m_{\rm BH} > 10^{-4}\,M_\odot\,{\rm yr}^{-1}$, to match the approximate
AGN detection threshold  in the LIRG sample. With this limit  the
IR-bright galaxy fraction in the 
RSA sample of Seyfert galaxies is then $\sim 22\% \pm 5\%$. 

Figure~\ref{fig:SFRoverBHAR_onlyRSA}
shows the distributions of the nuclear and total SFR/$\dot m_{\rm BH}$
for the all the RSA 
Seyferts indicating which ones are also IR-bright. Clearly, those RSA Seyfert
classified as IR-bright have ratios similar to, although slightly smaller
than,  those of the complete sample of local LIRGs. 
Given the fact that 
the AGNs in the RSA sample are not very different from those in
the LIRGs, the results above would suggest
that the bright AGN phase comes after and is somewhat distinct from the 
LIRG star forming phase. 
This interpretation would be
 consistent with observational works
showing a delay between the onset of the star formation activity and
the latter feeding of the AGN with the consequent BH growth 
\citep{Davies2007,  Wild2010} as well as with predictions 
from numerical simulations \citep[e.g.,][]{Hopkins2012}.

\subsection{SFRs versus AGN Detection in local LIRGs}
It is well known that the AGN
fraction increases with increasing IR 
luminosity from LIRGs to ULIRGs, and at the highest IR luminosities the
AGN might dominate bolometrically the luminosity of the system
\citep{Veilleux1995, Veilleux2009,Yuan2010,Nardini2010,AAH12}.  
We now examine further the possibility that the
IR-bright AGN phase 
comes after the distinct IR-bright star-forming phase. If that were the case, we
would expect the AGN fraction in LIRGs to be higher at the lowest
SFR. The AGN fraction here is computed only for those LIRGs
with a secure AGN detection, that is, a Seyfert classification
and/or the presence of mid-IR [Ne\,{\sc v}] lines. As in the previous
section, we computed the integrated SFR of the individual galaxies
from $L_{\rm IR}$ after subtracting the AGN contribution.

\begin{figure}
\includegraphics[width=0.25\textwidth,angle=-90]{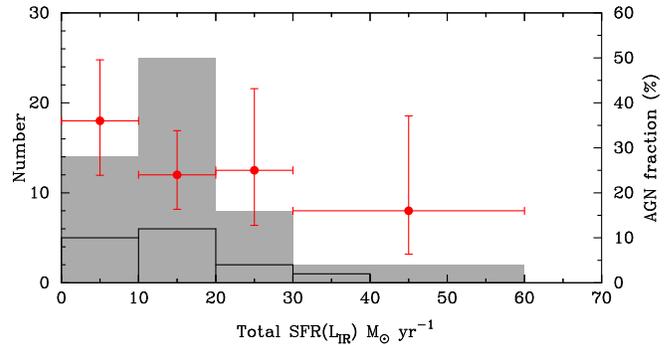}

\caption{Distributions of the integrated SFRs of the local LIRGs (grey
  histograms) and of those with a Seyfert classification and/or [Ne\,{\sc
  v}] emitters (solid line
  histograms). The data points 
  represent the AGN fraction (scale on the right hand vertical axis)
  in SFR intervals with $1\sigma$ error bars for the AGN fraction.}
\vspace{0.5cm}
\label{fig:SFRwithAGNfraction}
\end{figure}

Figure~\ref{fig:SFRwithAGNfraction} shows the AGN fraction as a
function of the total SFR of the galaxy. There is a tendency for LIRGs
in lowest SFR bin to have a higher AGN incidence ($36\%^{+14\%}_{-12\%}$) than
those in the highest SFR bin  ($17\%^{+20\%}_{-11\%}$).
Alternatively, this could be explained if it is harder to detect AGN 
in LIRGs with very high
SFRs, especially if it takes place mostly in the nuclear regions. 
Conversely, it may be easier to identify an AGN for low nuclear SFRs and
extended star formation. The
first explanation is not likely however, as most of the LIRGs in our sample in
the high SFR bin
also tend to have rather extended SF.  In other words, from
Figure~\ref{fig:SFRs} 
(right panel) there is no tendency for the fraction of nuclear
SFR to increase with the total SFR of the galaxy. In fact, in local
LIRGs the mid-IR emission  (both from AGN and star formation activity)
only starts to be highly concentrated in the
central regions for IR luminosities above $10^{11.8}\,L_\odot$, as
shown by \cite{DiazSantos2010}. We  conclude that the higher AGN
incidence at low SFRs in local LIRGs provides further evidence for a
time delay between the peaks of the star formation and BH growth.

\subsection{Key differences between merger and non-merger local LIRGs}
Before we explore the role of the LIRG phase in the context of star
formation activity and BH growth of massive late-type galaxies (next
section), it is  
important to point out the differences in the star formation histories
of merger and non-merger LIRGs 
in the local universe.

The triggering of the activity in local ULIRGs and likely also in the
most luminous local LIRGs is driven by 
major mergers. As a consequence, ULIRGs tend to host more luminous AGN
(quasar or nearly quasar-like luminosities), 
slightly more massive BH and have higher AGN bolometric contributions
that local LIRGs \citep[see Section~\ref{s:resolvedlines}
and][]{Veilleux2009, Nardini2010, 
AAH12}. Local merger LIRGs 
do show evidence  for a recent and intense period of 
star formation that consumed the gas faster, that is, a {\it
  bursty} SFR \citep[see e.g.,][]{AAH00, AAH01} 
as in local ULIRGs \citep[][]{RodriguezZaurin2010}. 
In this respect, 
most of the local merger-LIRGs could be considered as sub-ULIRGs and 
they might go  through a future ULIRG phase or may already have
experienced such a 
phase  \citep{Murphy2001}. 

Major mergers, however,  do not dominate in
numbers the population of local LIRGs,  especially 
at $L_{\rm IR} < 3\times 10^{11}\,L_\odot$, as inferred from  
morphological studies
\citep[see e.g.,][]{Sanders2004,AAH06,Kaviraj2009}. 
Therefore, in most local LIRGs the current episodes of star
formation activity and BH growth are likely the result of a less {\it
  violent} process, e.g., via minor 
mergers, fly-by companions, and/or secular evolution. 
 \cite{Poggianti2000} showed that most isolated galaxies in
their sample of IR-bright galaxies showed on average more moderate
optical Balmer absorption features than the strongly interacting systems in
their sample. This suggests that in non-merger LIRGs there is
not a strong contribution from a dominant  post-starburst stellar
population that would be associated with a recent bursty star
formation history.  This agrees with the finding that most of the 
mass  in non-merger LIRGs is from  evolved stellar
populations formed a few Gyr  
ago with a much smaller contribution in mass from a 
relatively young stellar population  \citep{Kaviraj2009, AAH06, AAH10}.

\subsection{The role of the LIRG-phase in the growth of BH at 
$L_{\rm IR} < 3\times
10^{11}\,L_\odot$}\label{sec:interpretation}

\begin{figure}
\vspace{-2cm}

\hspace{-1.5cm}
\includegraphics[width=0.45\textwidth,angle=-90]{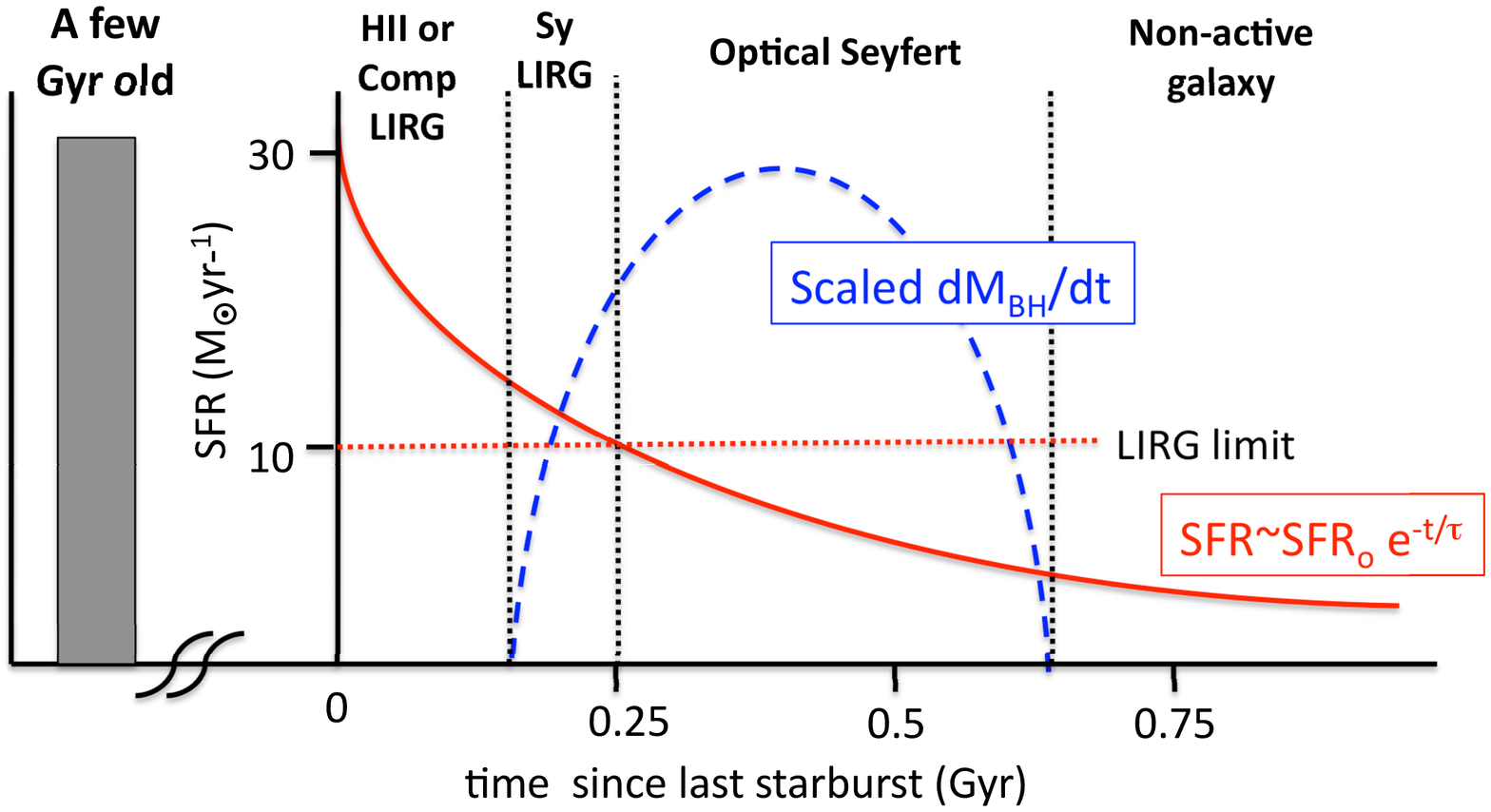}
\caption{Cartoon (not drawn to scale) depicting 
the proposed scenario the current episode of star formation and BH
growth of a 
  local non-major merger LIRG with $L_{\rm IR} < 3\times
  10^{11}\,L_\odot$. The last episode of SF is assumed to be 
an exponentially decaying SFR (red curve) with an e-folding time of
$\tau=2.5\times 10^8\,{\rm yr}$. The current period of BH growth is
assumed to have a  $5\times 10^{8}\,{\rm yr}$ lifetime (blue curve),
as in local galaxies with $M_{\rm BH}< 10^8\,M_\odot$
\citep{Marconi2004}. The BHAR would be on average $\sim 3000$
times lower than the SFR during the IR-bright phase.}.
\label{fig:cartoon}
\end{figure}

The cartoon in
Figure~\ref{fig:cartoon} illustrates our proposed scenario for the
current episode of BH
growth in a non-merger local LIRG typically with $L_{\rm IR} < 3\times
10^{11}\,L_\odot$ (i.e., integrated SFR$<30\,M_\odot \,{\rm yr}^{-1}$). This IR luminosity marks approximately the border
between local populations of LIRGs dominated by major merger and populations
of disky and relatively isolated LIRGs \citep{Sanders2004}.

We assume that the galaxy has an old stellar
  population formed  a few Gyr ago  \citep{Kaviraj2009,AAH10} and an
  existing BH.
Since the galaxy did not experience a recent
  major merger but rather a less {\it violent} process that induced the
  last episode of SF, we can assume an
  exponentially decaying SFR for it, as ${\rm SFR}(t) =
  {\rm SFR}_0\, {\rm e}^{-t/\tau}$ where $\tau$ is the
  e-folding time. The peak ${\rm SFR}(t=0) \simeq 30\,M_\odot\,{\rm
    yr}^{-1}$ is driven by the $L_{\rm IR}$ limit set 
above. From the position of local AGN on the
  color-magnitude diagram, \cite{Schawinski2009} demonstrated that the
  e-folding time cannot be too short (a few Myr or bursty) or 
too long (i.e., constant SFR). The high AGN detection rate in local LIRGs
\citep{Veilleux1995,Yuan2010,AAH12} implies that LIRGs are able to feed the BH
(i.e., show AGN activity)  roughly during half of the LIRG-phase (defined as
having integrated 
SFR$>10\,M_\odot\,{\rm yr}^{-1}$, see Figure~\ref{fig:SFRs}).
 Moreover, given the moderate BH masses inferred for LIRGs
 (Section~\ref{s:resolvedlines}), this AGN activity is likely to be 
sustained for 
$\sim 5 \times 10^8\,{\rm yr}$, that is, the typical lifetime of the BH growth for 
$M_{\rm BH}<10^8\,M_\odot$ \citep[see][]{Marconi2004}.

In this simple scenario, the galaxy would be
first classified optically  as an H\,{\sc ii} LIRG and would show elevated 
integrated and nuclear SFRs. Provided that there is a mechanism to
bring the gas  near the BH gravitational potential, after a time the existing BH starts accreting material 
at a sufficiently high rate.  Then the galaxy would still have a Seyfert-like
AGN luminosity.  Because there is an elevated nuclear SFR, the galaxy
would be optically classified first as composite (AGN/SB) and later on 
as a Seyfert LIRG as the SFR decreases. In both
cases the SFR/$\dot m_{\rm BH}$ ratios would be on average 
higher than those of optically
identified Seyferts.  After a few
hundred million years the SFR would be 
below $10\,M_\odot\,{\rm yr}^{-1}$ and thus it would not be 
classified as a LIRG any longer. However, the AGN phase may be
still active for another few hundred million years as Seyfert-like AGNs 
do not require large amounts of gas
to remain active \citep{Hopkins2012}. At this point 
the galaxy would be identified as an optical
Seyfert with a lower value of the SFR/BHAR ratio and a moderate SFR.
Eventually both the SF and the BHAR would be low and the galaxy may
not be identified as a Seyfert. 
A SFR e-folding time of $\tau \sim 2.5\times 10^8\,$yr
and a BH growth lifetime  of 
$\sim 5\times 10^8\,{\rm yr}$ would reproduce 
the SFRs  as well as the Seyfert and
composite fractions of local LIRGs \citep{Yuan2010}. 

In summary, in the local universe LIRGs with 
$L_{\rm IR}<3\times 10^{11}\,L_\odot$ LIRGs and an AGN
represent an early phase of 
the (possibly episodic) growth of BHs in massive spiral 
galaxies with high SFR, not necessarily
associated with a major merger event. An H\,{\sc ii}-like LIRG phase
would predate the AGN classified LIRG phase. The latter in turn would be
followed by an optically identified AGN phase where the galaxy still
shows an appreciable 
SFR but below the LIRG limit. Although the triggering mechanisms for
the SB and AGN activity of LIRGs and ULIRGs might be different, 
it appears that the sequence of events is similar. Namely, ULIRGs are
believed to involve the interaction
between two gas-rich galaxies that results first in a high SFR and then  an 
optical QSO phase  \citep[see][and references therein]{Sanders1988, Sanders1996,
  Yuan2010}. The end product, in contrast, is rather different with 
ULIRGs probably evolving into  moderate-mass ellipticals
\citep{Sanders1996,Dasyra2006,Hopkins2008} and non-merger 
LIRGs likely remaining late-type galaxies.

\section{Summary}
We have studied the current co-evolution of the star formation activity
and BH growth for a complete sample 
of 51 local LIRGs \citep[see][for details on the
sample]{AAH06,AAH12}.  We used {\it Spitzer}/IRS SH ($R\sim 600$) 
data in the $\sim 10-18\,\mu$m spectral
range to estimate BH masses from 
resolved \Neiii \, fine structure lines, and to measure  nuclear SFRs
from the $11.3\,\mu$m PAH feature and the \Neii \, line.

We detected spectrally resolved \Neiii \, lines in 19 LIRGs with gas
velocity dispersions  between 102 and $368\,{\rm km\,s}^{-1}$. Most
of these are LIRGs classified as Seyferts and/or are [Ne\,{\sc v}]
emitters. We assumed that \Neiii \, probes gas arising from 
the NLR in the AGN LIRGs and then used the
\cite{Dasyra2011} relation between $M_{\rm BH}$ and \Neiii \, velocity
dispersion. We combined gas velocity dispersions from \Neiii \, and  [O\,{\sc
  iii}]$\lambda$5007\AA \, and stellar velocity dispersions from the
literature to obtain BH masses 
 between $M_{\rm BH}=6\times 10^{6}\,M_\odot$ and 
$M_{\rm BH}=3.5\times 10^{8}\,M_\odot$ (median value of 
$3\times10^{7}\,M_\odot$). The derived AGN bolometric luminosities of
$L_{\rm bol}({\rm AGN}) =
(0.4-50)\times 
10^{43}\,{\rm erg\,s}^{-1}$ \citep[median of 
$\sim 1.4 \times 10^{43}\,{\rm erg \,s}^{-1}$, see ][]{AAH12} 
imply BHAR of $\dot m_{\rm BH} \simeq 0.0007 - 0.08\,M_\odot\,{\rm
  yr}^{-1}$  and relatively low  
Eddington ratios (typically $2\times 10^{-2}$).

Although AGNs accompany the star formation activity in the large majority
of local LIRGs \citep{Veilleux1995, Yuan2010, AAH12},  
in most cases the AGN is not energetically dominant \citep{AAH12}. 
Using the $11.3\,\mu$m PAH feature we obtained nuclear (on
typical scales of 1.5\,kpc) SFRs for the individual 
galaxies of between $\sim 0.8$ and $\sim 20\,{\rm
  M}_\odot \,{\rm yr}^{-1}$ and using $L_{\rm IR}$ integrated SFRs  between 
1 and $60\,M_\odot\,{\rm yr}^{-1}$  (for a Kroupa IMF). In a
large fraction of local LIRGs this 
on-going star formation activity is not only taking place in the
nuclear regions but it is also spread out on kpc 
scales throughout the galaxy. From the IR estimates in this work
the median value for the sample of local  LIRGs is
SFR(nuclear $\sim$ 1.5\,kpc)/SFR(total)$=0.5$. 

We found that local LIRGs have a time-averaged value of 
${\rm SFR}/\dot m_{\rm BH} \geq 2900$, which is a few times higher
than that of the
local population of bulge-dominated galaxies  
\citep[$\sim 10^3$, ][]{Heckman2004} and the 
local normalization of the BH mass versus bulge mass relation
\citep{Marconi2003, Haring2004}. However, this ratio is similar to the
value of local bulge galaxies averaged during the first $\sim
0.3\,$Gyr of the last starburst \citep{Wild2010}, that is, before the
peak of the BH growth.

Local LIRGs have on average higher ${\rm SFR}/\dot m_{\rm BH}$
ratios (both nuclear and total SFR) than
the optically selected RSA Seyferts  
in the local Universe. However, the IR-bright RSA Seyferts have
similar ratios to those of local LIRGs. Since the overall properties
of the AGNs in
local LIRGs and in the RSA sample are very similar, we concluded that 
the AGN phase comes after and is somewhat distinct from the LIRG star
forming phase. The AGN fraction in local LIRGs appear to be a function of the
integrated SFR, with LIRGs in the low SFR bin having a higher AGN (i.e.,
Seyfert) incidence than those with high SFRs. This provides further
support to the scenario that in local LIRGs the peak of intense IR-bright SF
activity occurs prior to
the AGN phase. 


Finally we put forward a simple model to explore the role of the LIRG-phase
in the last episode of BH growth  at $L_{\rm IR}<3\times
10^{11}\,L_\odot$. At these luminosities most 
LIRGs are not major-mergers, and have an old
stellar population formed a few Gyr ago and a current episode of
star formation. The latter only  a small mass fraction of the total
stellar mass.  We propose that 
an  exponentially decaying SFR with a peak ${\rm
  SFR}(t=0) \simeq 30\,M_\odot\,{\rm 
    yr}^{-1}$ and 
an e-folding time of $\tau \sim 2.5\times 10^8\,$yr together with 
a  $5\times 10^8\,$yr period  of  BH growth
would explain the observed SFRs and fractions of
Seyfert and composite objects in local LIRGs. In this scenario
non-merger LIRGs
hosting an AGN represent an early phase of the possibly episodic
growth of the BH which is followed by an optically Seyfert phase in massive
galaxies with relatively high SFR in their host galaxies. 
 
\section*{Acknowledgements}

A.A.-H. thanks the Astrophysics Department of the University of
Oxford, where most of this work was conducted, for their warm
hospitality. This work was supported in part by the Spanish Plan Nacional
de Astronom\'{\i}a y Astrof\'{\i}sica under grants AYA2009-05705-E and
AYA2010-21161-C02-1 and by the Augusto Gonz\'alez Linares Program
from the Universidad de Cantabria. 
M.P.-S. is funded by an ASI fellowship under contract I/005/11/0.
The authors wish to thank the referee for comments that helped
  improve the paper.

This research has made use of the NASA/IPAC Extragalactic Database
(NED) which is operated by the Jet Propulsion Laboratory, California
Institute of Technology, under contract with the National Aeronautics
and Space Administration.  This paper uses data products produced by
the OIR Telescope Data Center, supported by the Smithsonian
Astrophysical Observatory.

\end{document}